\documentclass[twocolumn, secnumarabic, amssymb, nobibnotes, prl, superscriptaddress]{revtex4}

\usepackage{graphicx} 
\usepackage{amsmath}
\usepackage{float}
\usepackage{color}
\usepackage[english]{babel}
\usepackage{array}
\usepackage[T1]{fontenc}
\usepackage[usenames,dvipsnames]{xcolor}
\usepackage{setspace}
\usepackage{hyperref}
\usepackage{bm}
\usepackage{lipsum} 
\usepackage{tikz}
\usepackage{stmaryrd}

\definecolor{color1}{rgb}{0.122,0.467,0.706}
\definecolor{color2}{rgb}{0.839,0.153,0.157}

\definecolor{color_RT1}{rgb}{1.0,0.455,0.0}
\definecolor{color_RT2}{rgb}{0.82,0.09,0.0}
\definecolor{color_RT3}{rgb}{0.545,0.0,0.0}
\definecolor{color_RT4}{rgb}{0.271,0.0,0.0}

\makeatletter
\renewcommand*{\fnum@figure}{{\normalfont \small{FIG.}~\thefigure}}
\makeatother

\begin{document}

\title{Memory from coupled instabilities in unfolded crumpled sheets}

\author{Dor Shohat}
\affiliation{Department of Condensed Matter, School of Physics and Astronomy, Tel Aviv University, Tel Aviv 69978, Israel}
\affiliation{Center for Physics and Chemistry of Living Systems, Tel-Aviv University, Tel Aviv 69978, Israel}

\author{Daniel Hexner}
\affiliation {Faculty of Mechanical Engineering, Technion, Haifa 32000, Israel}

\author{Yoav Lahini}
\affiliation {Department of Condensed Matter, School of Physics and Astronomy, Tel Aviv University, Tel Aviv 69978, Israel}
\affiliation{Center for Physics and Chemistry of Living Systems, Tel-Aviv University, Tel Aviv 69978, Israel}

\begin{abstract}
Crumpling an ordinary thin sheet transforms it into a structure with unusual mechanical behaviors, such as enhanced rigidity, emission of crackling noise, slow relaxations, and memory retention.  A central challenge in explaining these behaviours lies in understanding the contribution of the complex geometry of the sheet. Here, we combine cyclic driving protocols and 3D imaging to correlate the global mechanical response and the underlying geometric transformations in unfolded crumpled sheets. We find that their response to cyclic strain is intermittent, hysteretic, and encodes a memory of the largest applied compression. Using 3D imaging we show that these behaviours emerge due to an interplay between localized and interacting geometric instabilities in the sheet. A simple model confirms that these minimal ingredients are sufficient to explain the observed behaviors. Finally, we show that after training multiple memories can be encoded, a phenomenon known as return point memory. Our study lays the foundation for understanding the complex mechanics of crumpled sheets, and presents an experimental and theoretical framework for the study of memory formation in systems of interacting instabilities.
\end{abstract}

\maketitle

Before discarding a scrap piece of paper, we often crumple it offhandedly. Yet, the simplicity of this action stands in stark contrast to the complexity it introduces. A crumpled sheet has a wonderfully intricate geometry, and it exhibits unusual mechanical behaviors. These include enhanced rigidity \cite{matan2002crumpling,deboeuf2013comparative,croll2019compressive}, emission of crackling noise \cite{houle1996acoustic,kramer1996universal}, slow relaxations \cite{matan2002crumpling,lahini2017nonmonotonic}, and perhaps most staggeringly, various forms of memory. A crumpled sheet retains a memory of the shape it was deformed to \cite{kramer1996universal,oppenheimer2015shapeable}, a memory of the duration of past mechanical perturbations \cite{lahini2017nonmonotonic}, and a memory of the largest load it had been subjected to \cite{matan2002crumpling}. 

Understanding the emergence of these phenomena has been challenging, as it presumably depends on the complex structure as well as materials properties and plastic deformations they incur. While the roles of material aging and friction have been appreciated \cite{habibi2017effect,van2019tailoring,jules2020plasticity}, there is little insight into the role of geometry. Indeed, crumpling imprints in the sheet a disordered pattern of localized plastic deformations, created due to focusing of stress to point and line singularities known as d-cones and ridges \cite{cerda1999conical, witten2007stress}. Despite studies of the crumpling process \cite{aharoni2010direct,vliegenthart2006forced,gottesman2018state,andrejevic2021model} and the resulting three dimensional geometry \cite{PhysRevLett.57.791, blair2005geometry, PhysRevLett.96.136103, PhysRevE.81.061126,lin2009spontaneous,cambou2011three}, the relation between the complex geometry of a crumpled sheet and its unusual mechanics remains unexplored.

Here, motivated by recent studies on memory in matter \cite{keim2019memory}, we study the mechanical response of unfolded crumpled sheets under cyclic deformation. Such approach has been useful for characterizing the energy landscape \cite{corte2008random,keim2011generic,regev2021topology,mungan2019networks}, and identifying the effective degrees of freedom and their interactions \cite{sethna1993hysteresis,bense2021complex,keim2020global,mungan2019networks,van2021profusion} in several amorphous systems.  

We begin by studying the global mechanical response of elasto-plastic sheets that have been crumpled many times, and then opened to a rather flat configuration. We find that under cyclic strain, the force-displacement relations form hysteresis loops decorated with a myriad of abrupt force jumps. After a small number of cycles these converge to approximate limit cycles with repeating features. We show that this response encodes a memory of the largest applied compression, reminiscent of the memory of largest load reported in \cite{matan2002crumpling}. Interestingly, we find that afterreaching a limit cycle, the sheets exhibit return point memory \cite{keim2019memory} with high accuracy.

Next, we correlate the global response to local transformations in the sheets' geometry. We find that each force jump in the response corresponds to the snapping of a localized bistable degree of freedom - a \textit{Hysteron}. Each hysteron forms a two state hysteresis loop, and can be seen as a mechanical "bit". We measure both positive (ferromagnetic) and negative (anti-ferromagnetic) interactions between neighboring hysterons. Altogether, many such hysterons comprise the global mechanical response. Their collective states and transitions constitute the observed memory effects. Simulations of a network of interacting hysterons reproduce the observed behaviors, and highlight the inherent frustration. Our picture forms a basis for studying the mechanics of spatially coupled bistable elements \cite{lindeman2021multiple,bense2021complex,van2021profusion,keim2021multiperiodic}.

\section*{Results and Discussion}

\subsection*{Global Response - Memory and Intermittency}
 
A thin Mylar sheet is crumpled several times and then opened to an approximately flat configuration. In contrast to its floppy, uncrumpled ancestor, the crumpled-then-flattened sheet is stiffer and can easily carry its own weight if held at one end. When forcefully bent the sheet yields, but via a series of discrete jumps accompanied by audible popping sounds \cite{houle1996acoustic,kramer1996universal}. If the force is removed, the crumpled sheet tends to retain its bent shape, and to resist reversing the deformation \cite{kramer1996universal, oppenheimer2015shapeable}.
 
 \begin{figure*}[t]
\centering
\includegraphics[width=0.9\textwidth]{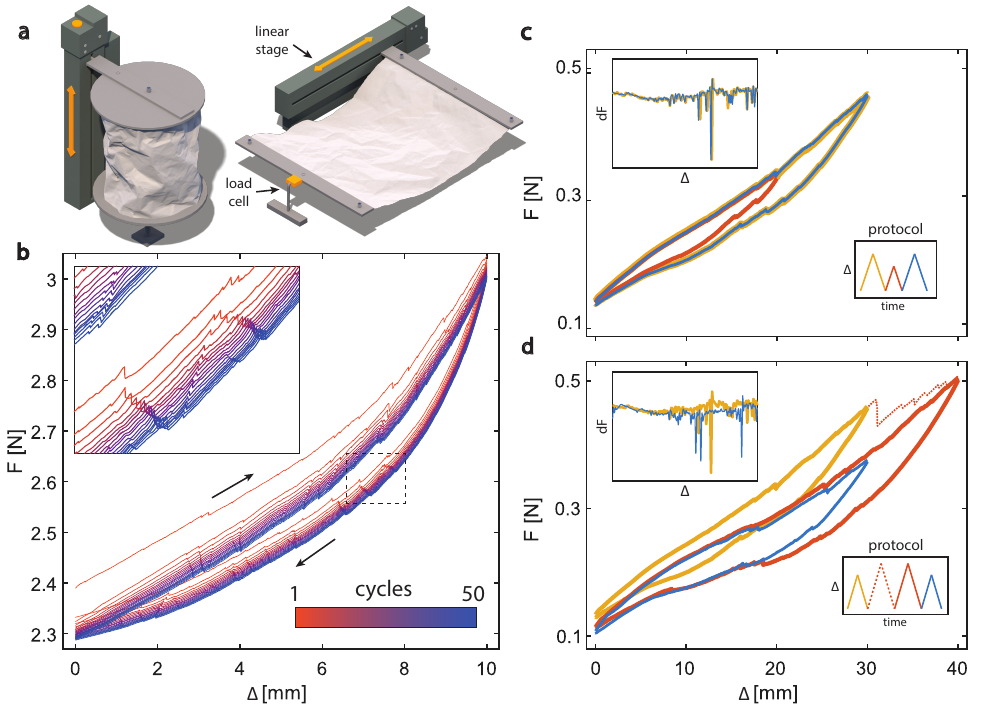}
\vspace*{-0.3cm}
\caption{\textbf{Global response} - ($\textbf{a}$) Illustration of experimental setup: two custom mechanical testers with sheets held at both ends in cylindrical (left) and flat (right) geometries.  ($\textbf{b}$) The force as a function of the displacement under oscillatory drive (every $3^{rd}$ cycle is shown). The curves converge to approximate limit cycles with repeating features. ($\textbf{c}$) Reversible behaviour, occurs when lowering and then restoring the initial displacement amplitude. ($\textbf{d}$) Irreversible behaviour occurs when the displacement surpasses previous maximal value, indicating a memory of largest displacement. Bottom right insets show the driving protocols. Top left insets show the force derivative $dF$ during compression. The sharp spikes allow to identify the instabilities. Note  that  the  pattern of spikes changes  in  the  irreversible case. }\label{fig1}
\end{figure*}

To measure this response we load the sheets into two custom mechanical testers, shown in Fig. \ref{fig1}a. The sheets are held at two opposite edges, where one edge is fixed and the displacement of the other is controlled by a linear motorized stage. The force exerted back by the sheet is measured using a load cell connected to the fixed end (see Methods). The rather flattened configuration of the sheets prevents friction from playing a role, as opposed to 3D configurations \cite{van2019tailoring}. We note, that in all experiments below, the measured stresses are much smaller than those applied during crumpling, therefore we expect negligible generation of new plastic deformations \cite{andrejevic2021model}.

We begin by considering the mechanical response of a sheet that is strained periodically. We vary the stage displacement $\Delta$ at a slow constant rate between $\Delta_{0}^{min}$ and $\Delta_{0}^{max}$. A typical example of the resulting force-displacement curve is presented in Fig. \ref{fig1}b. Two prominent features are observed. First, the mechanical response is hysteretic -- the force depends on the direction of change in the displacement. Secondly, the force curves are not smooth but rather decorated with  a multitude of sudden force jumps. After a small number of cycles the hysteresis curves converge to an approximate limit cycle, in which force jumps occur at nearly the same displacement values. Over many cycles we observe slow creep: a sluggish decrease in the measured force and a slow drift of the displacement values of the force jumps (see inset of Fig. \ref{fig1}b). We attribute these to material aging, presumably due to plastic flow in the creases \cite{jules2020plasticity}. Here we regard these creep processes as secondary effects and focus on the prominent, near-repeating features.

Next, we study the effect of changing the maximal displacement, after the system has reached a limit cycle in the interval $\left[\Delta_{0}^{min},\Delta_{0}^{max}\right]$. We find two distinct behaviours, depending on whether the new upper limit $\Delta_{1}^{max}$ is larger or smaller than the previous, $\Delta_{0}^{max}$.
For any $\Delta_{1}^{max}<\Delta_{0}^{max}$ the system immediately falls into a new limit cycle, in which the force-displacement curve is fully enclosed by (and partially overlaps with) the previous curve, as shown for example in Fig. \ref{fig1}c. This transition is reversible: upon returning to the initial $\Delta_{0}^{max}$ the previous limit cycle is immediately recovered, identified by the same pattern of force jumps (inset of Fig. \ref{fig1}c).

In contrast, for any $\Delta_{1}^{max}>\Delta_{0}^{max}$, the force-displacement curve changes irreversibly, as shown for example in Fig. \ref{fig1}d. After a short transient, the system converges to a new limit cycle, which generally has small overlap with the previous curve. Moreover, returning to $\Delta_{0}^{max}$, does not recover the original limit cycle. The new limit cycle is tilted with respect to the previous one, and exhibits a different pattern of force jumps (see inset of Fig. \ref{fig1}d). This indicates that the response is history dependent, and encodes a memory of the largest applied strain.

\subsection*{Return Point Memory}

Having shown that when completing a cycle in the reversible regime the system recovers it initial state, we investigate whether more complex displacement cycles obey the same rule. The phenomenon where the system returns to its initial state irrespective of the displacement trajectory, given it is confined to a previously reached limit cycle, is known as return point memory (RPM) \cite{keim2019memory,sethna1993hysteresis,deutsch2004return}. Systems exhibiting RPM can be used to encode and retrieve multiple memories repeatedly \cite{keim2019memory,perkovic1997improved}.

\begin{figure}
\centering
 \vspace*{-0.5cm}
\includegraphics[width=0.4\textwidth]{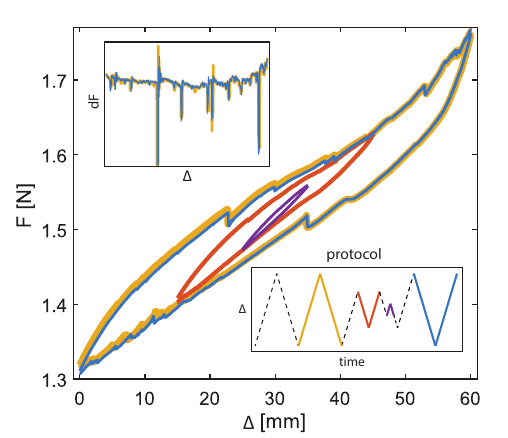}
\caption{\textbf{Return point memory} - A set of nested hysteresis loops indicating return point memory. Top left inset shows the force derivative $dF$ during compression. Bottom right inset shows the displacement sequence. }\label{fig2}
\end{figure}

One of the hallmarks of RPM is a hierarchy of nested hysteresis loops. Each time the displacement returns to previously visited value it recovers its previous state, closing a hysteresis loop. As a result, a displacement sequence that backtracks and closes smaller inner loops, results in a set of hysteresis loops that are confined within each other.  The inset of Fig. \ref{fig2} shows the displacement sequence that forms three loops (indicated in yellow, orange and purple). The resulting force-displacement curve yields three nested hysteresis loops in agreement with the occurrence of RPM. Finally, repeating a cycle in the parent interval (shown in blue)  yields a hysteresis loop that overlaps the initial cycle  (yellow) with a matching pattern of force-jumps. This amounts to strong evidence that return point memory is present.

\subsection*{Mesoscopic Excitations}

\begin{figure}[t]
\centering
\includegraphics[width=0.45\textwidth]{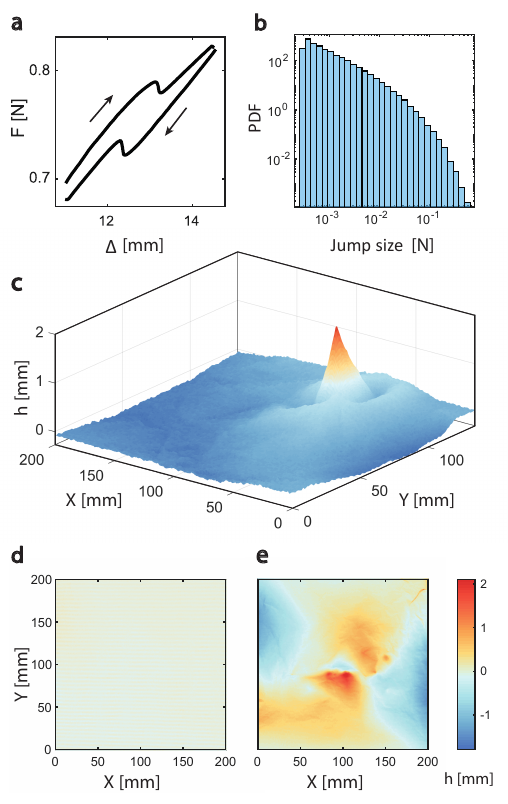}
\caption{\textbf{Mesoscopic excitations} - (\textbf{a}) {Experimental measurement of a single hysteretic event, demonstrates bistability. (\textbf{b}) Distribution of the magnitude of the force jumps. (\textbf{c}) Difference in the whole sheet's topographies following a force jump, measured through 3D scanning. Note the changes are localized to a small region in the sheet. 
(\textbf{d}) Negligible change in the sheet's topography following the sequence of cycles shown in the inset of Fig. \ref{fig1}c, leading to reversible behaviour. This indicates that by the end of the protocol all hysterons returned to their initial state. 
(\textbf{e}) Significant changes in sheet's topography following the sequence of cycles shown in the inset of Fig. \ref{fig1}d, leading to irreversible behaviour. This indicates that by the end of the protocol some of the hysterons changed their state irreversibly.}}
\label{fig3}
\end{figure}

We now turn to investigate the mechanism underlying the mechanical and memory response reported above. Since the force jumps are a central feature of the mechanical response, we start by investigating their geometric origin. We characterise a single force jump by varying the motor displacement periodically in a small interval around one such event. The resulting force-displacement curve shows that each event is reversible. An example of one such measurement is shown in Fig. \ref{fig3}a. However, the threshold for the transition in the increasing strain direction is larger than the threshold required to reverse the transition. Thus, a single event defines a two-state hysteresis loop, a \textit{hysteron}. For each hysteron we denote the flipping thresholds $\Delta_{i}^{+}>\Delta_{i}^{-}$. Collecting the magnitudes of many force jumps in the increasing displacement phase of the cycle, we find their distribution is broad, spanning several decades in magnitude (see Fig. \ref{fig3}b).

The planar geometry of our crumpled sheets enables direct imaging of the geometric transformations occurring at an instability. Using a 3D scanner, we measure the topography of the sheet before and after a single force jump. Subtracting the two measurements reveals that each force jump is a result of a localized event, in which a small region of the sheet snaps suddenly, in a direction perpendicular to the plane of the sheet. An example is shown in Fig. \ref{fig3}c. The bistable geometric features are localized, and for the most part consist of tips of d-cones, vertices formed at the crossing points of ridges, or small flat facets that undergo buckling \cite{kramer1996universal,lechenault2015generic}. We note that the instabilities are accompanied by audible acoustic emissions, and the resulting crackling noise was shown to exhibit a universal power-law distribution of intensities \cite{houle1996acoustic,kramer1996universal,sethna2001crackling}.

To relate this mesoscopic picture to the macroscopic memory of largest strain, we cycle the displacement between $\left[\Delta_{0}^{min},\Delta_{0}^{max}\right]$, and compare the topography of the sheet at the end of the cycle to the topography obtained after varying the upper limit $\Delta_{1}^{max}$. For the reversible regime of Fig. \ref{fig1}c, we find no discernible difference between the topographies (Fig. \ref{fig3}d). Namely, by the end of the protocol all hysterons return to their original state. This holds for any $\Delta_{1}^{max}<\Delta_{0}^{max}$.  In contrast, for the irreversible transition of Fig. \ref{fig1}d, the topography changes significantly as well (see Fig. \ref{fig3}d). In particular, several hysterons have changed their state. This is generally true for any $\Delta_{1}^{max}>\Delta_{0}^{max}$. We therefore deduce that the memory of largest strain is encoded in the collective configuration of the hysterons.

We note, that this mechanism  may also play a role in the previously observed memory of the largest applied load \cite{matan2002crumpling}. However, in those experiments, the formation of new creases could also contribute to the reported memory. Presumably, new creases were formed due to the confined geometry and the large loads \cite{gottesman2018state}.

\subsection*{Interactions Between Hysterons}

To complete the mesoscopic description, we ask whether interactions between the bistable elements are present and affect the sheet's mechanical response. Interactions between hysterons are expected to be mediated through the sheet's elasticity; flipping a hysteron presumably modifies the local strain field, affecting the flipping thresholds of its neighbors \cite{oppenheimer2015shapeable,plummer2020buckling}. Thus, interactions may be inferred from shifts in the activation thresholds which are configuration dependent \cite{bense2021complex,van2021profusion,lindeman2021multiple,mungan2019networks}. In experiments, we find evidence for such interactions both by probing flipping thresholds locally, and by carefully examining the global response of the sheet.

To probe interactions locally, one must measure how the flipping threshold of a hysteron depends on the state of a neighboring one. However, the experimental setup of Fig. \ref{fig1}a measures in-plane forces, and does not allow isolating the response of a single hysteron. To this end, we attach a probe to a load cell and use it to push locally on a hysteron in a direction perpendicular to the sheet (out of plane), until it flips. We control the displacement of the probe $\delta$ while measuring the normal force. The measured force first increases, until an instability occurs - the hysteron flips and the force drops to zero. This instability identifies the local flipping threshold, denoted $\delta_{i}^{+}$. 

\begin{figure}
\centering
 \vspace*{-0.5cm}
\includegraphics[width=0.48\textwidth]{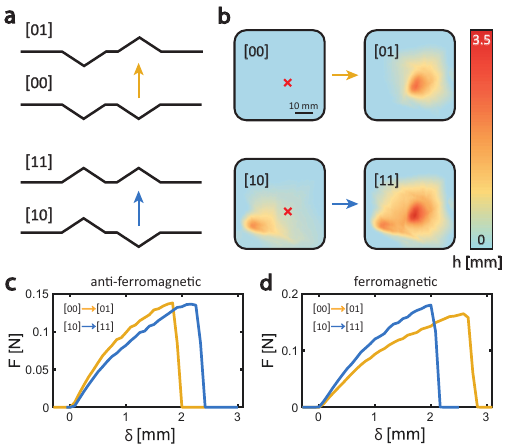}
\caption{\textbf{Local probing of interactions between hysterons} (\textbf{a}) Illustrations of the states of two neighboring hysterons. A probe connected to a load cell pushes the right hysteron from below, while measuring the transitions $\left[00\right] \rightarrow \left[01\right]$ and $\left[10\right] \rightarrow \left[11\right]$, yellow and blue arrows respectively. (\textbf{b}) Example for experimental state measurements: we obtain the 3D topographies of each state, and substract the topography of $\left[00\right]$. The difference height maps are focused to a 5$cm$ by 5$cm$ region of the sheet and reveal the localized hysterons. The red cross marks the location of the load cell's force measurements. (\textbf{c}-\textbf{d}) Force-displacement measurements of the local probe, where sharp drops correspond to the flipping instability. Panel \textbf{c} shows an example for anti-ferromagnrtic interactions; The flipping threshold for the right hysteron increases when the left hysteron is flipped up. This example corresponds to the topographic maps of panel \textbf{b}. Panel \textbf{d} shows an example of ferromagnrtic interactions in a different realization (i.e., for another pair of hysterons); the flipping threshold for the right hysteron decreases when the left hysteron is flipped up.}\label{fig_int}
\end{figure}

We identify two neighboring hysterons in the sheet that are flipped down, and denote their state $\left[00\right]$. We then measure the flipping threshold for the first hysteron $\delta_{1}^{+}$, and its dependence on the state of the second hysteron. We compare the transitions $\left[00\right] \rightarrow \left[01\right]$ and $\left[10\right] \rightarrow \left[11\right]$, as shown in Fig. \ref{fig_int}a,b. Doing so for several pairs of hysterons we find that their flipping thresholds are state dependent. We find both anti-ferromagnetic and ferromagnetic interactions: the flipping threshold of a hysteron may increase or decrease depending on the state of its neighbor. Examples for both types of interactions are shown in Fig. \ref{fig_int}c,d respectively. We also observe cases in which locally flipping a hysteron results in the spontaneous flip of another one in a small avalanche.
 Overall, this amounts to evidence for significant interactions between hysterons in the sheet.

Interactions between hysterons are also apparent in the global response of the sheet to strain applied at the boundaries, as reported above. This is most clear in Fig. \ref{fig1}d, when comparing the two limit cycles before and after surpassing the maximal displacement (yellow and blue curves respectively). Even though the displacement range is identical for the two cycles, the pattern of instabilities changes considerably. This is identified from the spikes in the derivative of the force, shown in the inset of Fig. \ref{fig1}d. When the maximal amplitude is exceeded (red curve), newly activated hysterons change the strain field across the sheet, affecting the flipping thresholds of their neighbors and therefore changing the pattern of instabilities. In contrast, if interactions had not been present, instabilities would occur at fixed displacements set only by their individual thresholds \cite{preisach1935magnetische}.

\begin{figure*}
\centering
\includegraphics[width=1\textwidth]{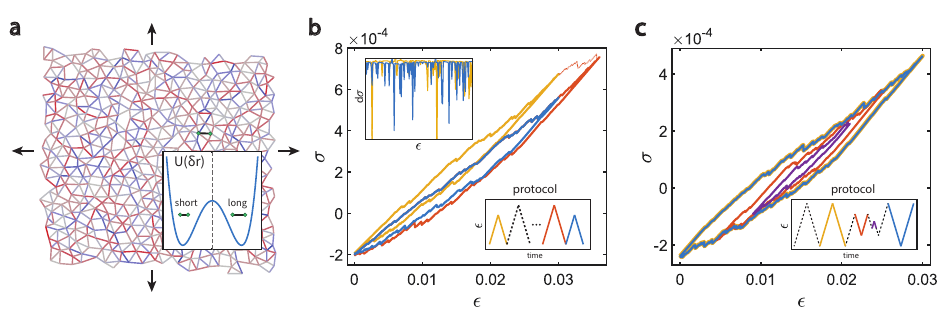}
\vspace*{-0.8cm}
\caption{\textbf{Coupled hysterons model} - (\textbf{a}) A disordered bonded network where each bond is bistable at zero strain. The bond's double minima potential is shown in the inset. In simulation the network is strained isotropically and quasistatically. Bond colors represent excess stresses, tension (blue) and compression (red). (\textbf{b}) Surpassing the maximal the strain alters the stress-strain relation at smaller strain amplitudes, similarly to the experiments in Fig. \ref{fig1}d. Note the tilt in the new curve. The top left inset shows the stress derivative, revealing  a new pattern of instabilities. (\textbf{c}) Nested hysteresis loops indicate return point memory, similarly to Fig. \ref{fig2}. Bottom insets of (\textbf{b}) and (\textbf{c}) show the strain sequence. The network was initially trained over 10 cycles.}
\label{fig5}
\end{figure*}

Another intriguing effect seen after surpassing the maximal displacement is a tilt of the force-displacement curve. After returning to the initial displacement interval, the force displacement curve (blue curve in Fig. \ref{fig1}d) is tilted with respect to the previous limit cycle (yellow). This again is not expected for a system of non-interacting hysterons \cite{mayergoyz1988generalized}, as we discuss in the supplementary material. This effect suggests that interactions allow the system to explore lower energy states (i.e., with smaller forces), thereby reducing its effective elastic modulus.

We note that the presence of anti-ferromagnetic interactions implies that crumpled sheets do not have a partial ordering of internal states. This property is termed \textit{no passing} (NP), and plays a key role in understanding RPM \cite{middleton1992asymptotic,sethna1993hysteresis,mungan2019structure}. While RPM has been observed in systems without NP such as various magnetic systems \cite{deutsch2004return,gilbert2015direct} and amorphous solids \cite{keim2020global,mungan2019networks}, its occurrence is not fully understood \cite{mungan2019structure}.

\subsection*{A Coupled Hysterons Model}

Our experiments have shown that the effective degrees of freedom of crumpled sheets are localized, coupled, bistable elements. To formulate a model we make further simplifying assumptions. We assume that the planar sheet can be approximated by a two dimensional system embedded with bistable elements. Though the majority of the motion in experiments occurs in the lateral dimension, we assume it yields an in-plane strain. Lastly, we incorporate interactions by assuming that the strain results in forces between the bistable elements. 

A simple realization of such a model is a two dimensional disordered bonded network, where each bond is a bistable spring \cite{kedia2019drive,benichou2013structures} (see Fig. \ref{fig5}a). Namely, each bond is governed by a potential with two minima, given by:
\begin{equation} 
\label{bistable}
U_i=\frac{c_i}{4}\left(\delta r_i\right)^{4}-\frac{a_i}{2}\left(\delta r_i\right)^{2}.
\end{equation}
Here, $\delta r_i$ denotes the distance from the local maximum. Each bond is defined by the two parameters $a_i$ and $c_i$, which set the locations of the two minima,  $\delta r_{min} = \pm \sqrt{a_i/c_i}$. The overall length of the bond is taken to be significantly larger than $\delta r_{min}$. Additional details are presented in the methods section and supplementary information. In the simulations, the network is strained isotropically and quasi-statically; at each step the energy is minimized to reach force balance. 

We measure the stress-strain relation and find smooth elastic intervals separated by sharp stress-drops. These instabilities originate from bonds transitioning between their two minima, as illustrated in the supplementary video. Under periodic drive the system may converge to a precise limit cycle, depending on the parameters of the model.  Even when it does not converge, often consecutive cycles yield very similar stress-strain curves, and appear as approximate limit cycles.

Following the same protocols as in experiments, we find that the model recovers all essential experimental findings:  memory of the largest strain, a tilt in the stress-strain curves and change to the activation pattern when increasing the strain amplitude, and, after training, the nested loops which characterize return point memory (Fig. \ref{fig5}b,c). Further characterization is provided in the supplementary information.

Interestingly, we observe that bonds generally deviate from their minima, due to a mismatch between their rest lengths. Namely, their geometrical constraints cannot all be satisfied simultaneously, and the network is geometrically frustrated \cite{moessner2006geometrical}. This is demonstrated in Fig. \ref{fig5}a and in the supplementary video, where residual stresses within the network are visible, even when the global stress vanishes. Such frustration also develops in networks which were initiated at an unfrustrated state. We expect such frustration to emerge in crumpled sheets as well, due to their disordered geometry and the anti-ferromagnetic interactions between hysterons. 

\subsection*{Conclusions \& Outlook}

We have studied the mechanical response of unfolded crumpled sheets to cyclic strain. We found that the force displacement curves are intermittent, hysteretic and encode a memory of the largest strain. All these effects can be traced back to the collective response of multiple bistable snap-through instabilities spread across the sheet. These mesoscopic, localized and interacting degrees of freedom form the basis for understanding the mechanics of crumpled sheets. A model of a coupled hysterons network reproduces the observed behaviors and sheds light on the role of interactions. Altogether, our work offers an experimental and theoretical framework for the study of memory formation in systems composed of interacting instabilities.

Remarkably, our system exhibits return point memory with high precision, allowing to encode multiple memories through a series of nested hysteresis loops. The occurrence of RPM is surprising given the complexity and strength of the interactions. Thus, a sheet that is crumpled mindlessly and with little effort is transformed into a programmable material, without the careful design required in engineered mechanical metamaterials \cite{chen2021reprogrammable,jules2022delicate,silverberg2014using,udani2021programmable,waitukaitis2015origami}. 

Further research is required to fully map the distribution of hysterons across the crumpled sheet, the patterns and range of their interactions, and how these are related to the geometry. Particular attention should be given to the formulation of the interactions between hysterons, which may not be of a pairwise nature, and may affect the activation thresholds $\Delta^{\pm}$ differently \cite{bense2021complex}.

The description of crumpled sheets as a frustrated network of coupled bistable degrees of freedom, suggests that they can be viewed as a mechanical spin glass with a complex energy landscape. Similar suggestions have been made in the context of origami \cite{stern2017complexity,chen2018branches,plummer2020buckling}. This begs the question whether the glass-like behaviors of crumpled sheets, such as logarithmic aging and Kovacs-like memory retention \cite{matan2002crumpling,lahini2017nonmonotonic}, could also be understood using this framework. Furthermore, as geometric frustration generically gives rise to an extensive multistability, this description may relate to the shape memory of crumpled sheets \cite{oppenheimer2015shapeable}. We hope our work promotes further understanding of these phenomena, both in crumpled sheets and in other systems that can be effectively described as a network of strongly coupled metastable elements.

\section*{Materials and Methods}

\subsection*{Experimental}
The experiments are performed on Mylar sheets, 25 and 90 microns thick and approximately 20$cm$ by 20$cm$ across. These were crumpled manually several times and then loaded the mechanical testers. The difference between the two setups is in the geometry of the testers' edge constraints. The first has straight constraints, keeping the sheet approximately flat and allowing direct imaging of its topography. However, in this setup the response is limited, as the sheet tends to bulge along the axis of strain, leaving the transverse dimension flat. The second cylindrical configuration avoids these localized deformations, as the transverse curvature frustrates the bending along the strain axis. In all the results displayed above, topographic 3D imaging is done with the flat setup using opaque 90$\mu m$ thick Mylar, and a HP David 5 3D scanner. Force-displacement curves are measured using the cylindrical setup and transparent 25$\mu m$ thick Mylar. We note that the flat setup exhibits the same characteristic features in it's response, including the memory of the largest strain, the signature of interactions, and return point memory.

\subsection*{Numerical model }
The picture suggested by the experiments is that a crumpled sheet
can be modeled as a disordered collection of coupled bistable mechanical degrees
of freedom. To test if these ingredients are sufficient to explain
the observed behaviors, we study a simple model. We assume that a
planar sheet can be approximated by a two dimensional elastic system
embedded with bistable elements. A simple realization of this is a
disordered bonded network where each bond is bistable, implemented
through a potential with two minima. The length of the bond is denoted
by $r=r_{0}+\delta r$, where $\delta r$ is smaller than $r_{0}$,
which signifies the average between the two bistable states. For simplicity
the potential is chosen to be symmetric : 
\[
U_{i}=\frac{C}{4}\left(\delta r\right)^{4}-\frac{a_{i}}{2}\left(\delta r\right)^{2}
\]
Here, $a_{i}>0$ and the two minima are at $\delta r=\pm\sqrt{a_{i}/c}$.
For simplicity, the constant $c$ is chosen to be the same for all
bonds. We consider two cases: (1) $a_{i}=a_{0}$ is the same for all
bonds. (2) $a_{i}$ is uniformly distributed between $\left[0,a_{0}\right]$.
To avoid boundary effects we employ periodic boundary conditions.

We simulate quasi-static dynamics, by straining the system between $0$ and $\epsilon_{max}$. The strain is discretized into small steps, below $10^{-4}$. Each step of the dynamics consists of a small change to the strain, and then an energy minimization step to reach force balance, using the FIRE algorithm \cite{FIRE}. The qualitative behavior appears to be independent of the particular deformation, and throughout the paper we strain both axes equally.

In Fig. 4a the number of nodes is  $N=512$. In Fig. 4b,c the number of nodes is  $N=2048$ ,  $a_i$ are uniformly distributed between $[0,0.04]$, $C=1$, and the average $<r_0>=1$.

\subsection*{Disordered network preparation}
We prepare our two dimensional network from amorphous packings of spheres at zero temperature \cite{OHern2003}. This choice is made for convenience and we believe has little impact on the results. A pair of spheres interact when the inter-particle distance, $r$, is below the sum of the radii $R_{i}+R_{j}$, via the potential:

\[
V_{ij}\left(r\right)=\begin{cases}
V_{0}\left(1-\frac{r}{R_{i}+R_{j}}\right)^{2} & r\leq R_{i}+R_{j}\\
0 & r>R_{i}+R_{j}
\end{cases}
\]
 
Force balance configurations are reached by minimizing the energy using the FIRE algorithm \cite{FIRE}. The properties of packings depends on the distance from the jamming transition, which can be tuned through the pressure exerted on the box. In the limit of zero applied pressure the system has critical like behavior, characterized by anomalous elasticity and diverging length scales. To avoid these atypical behavior we focus on the regime that is far from the jamming transition. The distance from the isostatic transition is often characterized by the excess coordination number $\Delta Z=Z-Z_{c}$. Here, $Z=2N_{b}/N$ where $N_{b}$ are the number of bonds, $N$ is the number of nodes and $Z_{c}\approx2d$ \cite{Phillips1979, alexander1998, moukarzel1998}. All our simulation are at $\Delta Z\approx1.5$, which far from the isostatic point $\Delta Z=0$. 

The packings are the converted to a bonded network, by identifying the nodes with the centers of spheres and connecting pairs of overlapping particles with a bond. The length of the bond $r_{0}$ is set by the distance between the particle pair. Initial we set $\delta r=0$, however, this is the local unstable maximum and the system relaxes into a stable basin of attraction.

\subsection*{Acknowledgements }
We are grateful to Martin van Hecke, Ido Regev, Naomi Oppenheimer, Yohai Bar Sinai, Roy Beck and Yair Shokef for enlightening discussions. We thank Lir Nizan for 3D graphic design. This work was supported by the Israel Science Foundation grant 2096/18 (YL) and grant 2385/20 (DH).
\bibliography{biblo}

\begin{thebibliography}{58}
\expandafter\ifx\csname natexlab\endcsname\relax\def\natexlab#1{#1}\fi
\expandafter\ifx\csname bibnamefont\endcsname\relax
  \def\bibnamefont#1{#1}\fi
\expandafter\ifx\csname bibfnamefont\endcsname\relax
  \def\bibfnamefont#1{#1}\fi
\expandafter\ifx\csname citenamefont\endcsname\relax
  \def\citenamefont#1{#1}\fi
\expandafter\ifx\csname url\endcsname\relax
  \def\url#1{\texttt{#1}}\fi
\expandafter\ifx\csname urlprefix\endcsname\relax\def\urlprefix{URL }\fi
\providecommand{\bibinfo}[2]{#2}
\providecommand{\eprint}[2][]{\url{#2}}

\bibitem[{\citenamefont{Matan et~al.}(2002)\citenamefont{Matan, Williams,
  Witten, and Nagel}}]{matan2002crumpling}
\bibinfo{author}{\bibfnamefont{K.}~\bibnamefont{Matan}},
  \bibinfo{author}{\bibfnamefont{R.~B.} \bibnamefont{Williams}},
  \bibinfo{author}{\bibfnamefont{T.~A.} \bibnamefont{Witten}},
  \bibnamefont{and} \bibinfo{author}{\bibfnamefont{S.~R.} \bibnamefont{Nagel}},
  \bibinfo{journal}{Physical Review Letters} \textbf{\bibinfo{volume}{88}},
  \bibinfo{pages}{076101} (\bibinfo{year}{2002}).

\bibitem[{\citenamefont{Deboeuf et~al.}(2013)\citenamefont{Deboeuf, Katzav,
  Boudaoud, Bonn, and Adda-Bedia}}]{deboeuf2013comparative}
\bibinfo{author}{\bibfnamefont{S.}~\bibnamefont{Deboeuf}},
  \bibinfo{author}{\bibfnamefont{E.}~\bibnamefont{Katzav}},
  \bibinfo{author}{\bibfnamefont{A.}~\bibnamefont{Boudaoud}},
  \bibinfo{author}{\bibfnamefont{D.}~\bibnamefont{Bonn}}, \bibnamefont{and}
  \bibinfo{author}{\bibfnamefont{M.}~\bibnamefont{Adda-Bedia}},
  \bibinfo{journal}{Physical review letters} \textbf{\bibinfo{volume}{110}},
  \bibinfo{pages}{104301} (\bibinfo{year}{2013}).

\bibitem[{\citenamefont{Croll et~al.}(2019)\citenamefont{Croll, Twohig, and
  Elder}}]{croll2019compressive}
\bibinfo{author}{\bibfnamefont{A.~B.} \bibnamefont{Croll}},
  \bibinfo{author}{\bibfnamefont{T.}~\bibnamefont{Twohig}}, \bibnamefont{and}
  \bibinfo{author}{\bibfnamefont{T.}~\bibnamefont{Elder}},
  \bibinfo{journal}{Nature communications} \textbf{\bibinfo{volume}{10}},
  \bibinfo{pages}{1} (\bibinfo{year}{2019}).

\bibitem[{\citenamefont{Houle and Sethna}(1996)}]{houle1996acoustic}
\bibinfo{author}{\bibfnamefont{P.~A.} \bibnamefont{Houle}} \bibnamefont{and}
  \bibinfo{author}{\bibfnamefont{J.~P.} \bibnamefont{Sethna}},
  \bibinfo{journal}{Physical Review E} \textbf{\bibinfo{volume}{54}},
  \bibinfo{pages}{278} (\bibinfo{year}{1996}).

\bibitem[{\citenamefont{Kramer and Lobkovsky}(1996)}]{kramer1996universal}
\bibinfo{author}{\bibfnamefont{E.~M.} \bibnamefont{Kramer}} \bibnamefont{and}
  \bibinfo{author}{\bibfnamefont{A.~E.} \bibnamefont{Lobkovsky}},
  \bibinfo{journal}{Physical Review E} \textbf{\bibinfo{volume}{53}},
  \bibinfo{pages}{1465} (\bibinfo{year}{1996}).

\bibitem[{\citenamefont{Lahini et~al.}(2017)\citenamefont{Lahini, Gottesman,
  Amir, and Rubinstein}}]{lahini2017nonmonotonic}
\bibinfo{author}{\bibfnamefont{Y.}~\bibnamefont{Lahini}},
  \bibinfo{author}{\bibfnamefont{O.}~\bibnamefont{Gottesman}},
  \bibinfo{author}{\bibfnamefont{A.}~\bibnamefont{Amir}}, \bibnamefont{and}
  \bibinfo{author}{\bibfnamefont{S.~M.} \bibnamefont{Rubinstein}},
  \bibinfo{journal}{Physical Review Letters} \textbf{\bibinfo{volume}{118}},
  \bibinfo{pages}{085501} (\bibinfo{year}{2017}).

\bibitem[{\citenamefont{Oppenheimer and
  Witten}(2015)}]{oppenheimer2015shapeable}
\bibinfo{author}{\bibfnamefont{N.}~\bibnamefont{Oppenheimer}} \bibnamefont{and}
  \bibinfo{author}{\bibfnamefont{T.~A.} \bibnamefont{Witten}},
  \bibinfo{journal}{Physical Review E} \textbf{\bibinfo{volume}{92}},
  \bibinfo{pages}{052401} (\bibinfo{year}{2015}).

\bibitem[{\citenamefont{Habibi et~al.}(2017)\citenamefont{Habibi, Adda-Bedia,
  and Bonn}}]{habibi2017effect}
\bibinfo{author}{\bibfnamefont{M.}~\bibnamefont{Habibi}},
  \bibinfo{author}{\bibfnamefont{M.}~\bibnamefont{Adda-Bedia}},
  \bibnamefont{and} \bibinfo{author}{\bibfnamefont{D.}~\bibnamefont{Bonn}},
  \bibinfo{journal}{Soft matter} \textbf{\bibinfo{volume}{13}},
  \bibinfo{pages}{4029} (\bibinfo{year}{2017}).

\bibitem[{\citenamefont{Van~Bruggen et~al.}(2019)\citenamefont{Van~Bruggen, Van
  Der~Linden, and Habibi}}]{van2019tailoring}
\bibinfo{author}{\bibfnamefont{E.}~\bibnamefont{Van~Bruggen}},
  \bibinfo{author}{\bibfnamefont{E.}~\bibnamefont{Van Der~Linden}},
  \bibnamefont{and} \bibinfo{author}{\bibfnamefont{M.}~\bibnamefont{Habibi}},
  \bibinfo{journal}{Soft Matter} \textbf{\bibinfo{volume}{15}},
  \bibinfo{pages}{1633} (\bibinfo{year}{2019}).

\bibitem[{\citenamefont{Jules et~al.}(2020)\citenamefont{Jules, Lechenault, and
  Adda-Bedia}}]{jules2020plasticity}
\bibinfo{author}{\bibfnamefont{T.}~\bibnamefont{Jules}},
  \bibinfo{author}{\bibfnamefont{F.}~\bibnamefont{Lechenault}},
  \bibnamefont{and}
  \bibinfo{author}{\bibfnamefont{M.}~\bibnamefont{Adda-Bedia}},
  \bibinfo{journal}{Physical Review E} \textbf{\bibinfo{volume}{102}},
  \bibinfo{pages}{033005} (\bibinfo{year}{2020}).

\bibitem[{\citenamefont{Cerda et~al.}(1999)\citenamefont{Cerda, Chaieb, Melo,
  and Mahadevan}}]{cerda1999conical}
\bibinfo{author}{\bibfnamefont{E.}~\bibnamefont{Cerda}},
  \bibinfo{author}{\bibfnamefont{S.}~\bibnamefont{Chaieb}},
  \bibinfo{author}{\bibfnamefont{F.}~\bibnamefont{Melo}}, \bibnamefont{and}
  \bibinfo{author}{\bibfnamefont{L.}~\bibnamefont{Mahadevan}},
  \bibinfo{journal}{Nature} \textbf{\bibinfo{volume}{401}}, \bibinfo{pages}{46}
  (\bibinfo{year}{1999}).

\bibitem[{\citenamefont{Witten}(2007)}]{witten2007stress}
\bibinfo{author}{\bibfnamefont{T.~A.} \bibnamefont{Witten}},
  \bibinfo{journal}{Reviews of Modern Physics} \textbf{\bibinfo{volume}{79}},
  \bibinfo{pages}{643} (\bibinfo{year}{2007}).

\bibitem[{\citenamefont{Aharoni and Sharon}(2010)}]{aharoni2010direct}
\bibinfo{author}{\bibfnamefont{H.}~\bibnamefont{Aharoni}} \bibnamefont{and}
  \bibinfo{author}{\bibfnamefont{E.}~\bibnamefont{Sharon}},
  \bibinfo{journal}{Nature Materials} \textbf{\bibinfo{volume}{9}},
  \bibinfo{pages}{993} (\bibinfo{year}{2010}).

\bibitem[{\citenamefont{Vliegenthart and
  Gompper}(2006)}]{vliegenthart2006forced}
\bibinfo{author}{\bibfnamefont{G.}~\bibnamefont{Vliegenthart}}
  \bibnamefont{and} \bibinfo{author}{\bibfnamefont{G.}~\bibnamefont{Gompper}},
  \bibinfo{journal}{Nature Materials} \textbf{\bibinfo{volume}{5}},
  \bibinfo{pages}{216} (\bibinfo{year}{2006}).

\bibitem[{\citenamefont{Gottesman et~al.}(2018)\citenamefont{Gottesman,
  Andrejevic, Rycroft, and Rubinstein}}]{gottesman2018state}
\bibinfo{author}{\bibfnamefont{O.}~\bibnamefont{Gottesman}},
  \bibinfo{author}{\bibfnamefont{J.}~\bibnamefont{Andrejevic}},
  \bibinfo{author}{\bibfnamefont{C.~H.} \bibnamefont{Rycroft}},
  \bibnamefont{and} \bibinfo{author}{\bibfnamefont{S.~M.}
  \bibnamefont{Rubinstein}}, \bibinfo{journal}{Communications Physics}
  \textbf{\bibinfo{volume}{1}}, \bibinfo{pages}{1} (\bibinfo{year}{2018}).

\bibitem[{\citenamefont{Andrejevic et~al.}(2021)\citenamefont{Andrejevic, Lee,
  Rubinstein, and Rycroft}}]{andrejevic2021model}
\bibinfo{author}{\bibfnamefont{J.}~\bibnamefont{Andrejevic}},
  \bibinfo{author}{\bibfnamefont{L.~M.} \bibnamefont{Lee}},
  \bibinfo{author}{\bibfnamefont{S.~M.} \bibnamefont{Rubinstein}},
  \bibnamefont{and} \bibinfo{author}{\bibfnamefont{C.~H.}
  \bibnamefont{Rycroft}}, \bibinfo{journal}{Nature Communications}
  \textbf{\bibinfo{volume}{12}}, \bibinfo{pages}{1} (\bibinfo{year}{2021}).

\bibitem[{\citenamefont{Kantor et~al.}(1986)\citenamefont{Kantor, Kardar, and
  Nelson}}]{PhysRevLett.57.791}
\bibinfo{author}{\bibfnamefont{Y.}~\bibnamefont{Kantor}},
  \bibinfo{author}{\bibfnamefont{M.}~\bibnamefont{Kardar}}, \bibnamefont{and}
  \bibinfo{author}{\bibfnamefont{D.~R.} \bibnamefont{Nelson}},
  \bibinfo{journal}{Physical Review Letters} \textbf{\bibinfo{volume}{57}},
  \bibinfo{pages}{791} (\bibinfo{year}{1986}).

\bibitem[{\citenamefont{Blair and Kudrolli}(2005)}]{blair2005geometry}
\bibinfo{author}{\bibfnamefont{D.~L.} \bibnamefont{Blair}} \bibnamefont{and}
  \bibinfo{author}{\bibfnamefont{A.}~\bibnamefont{Kudrolli}},
  \bibinfo{journal}{Physical Review Letters} \textbf{\bibinfo{volume}{94}},
  \bibinfo{pages}{166107} (\bibinfo{year}{2005}).

\bibitem[{\citenamefont{Sultan and Boudaoud}(2006)}]{PhysRevLett.96.136103}
\bibinfo{author}{\bibfnamefont{E.}~\bibnamefont{Sultan}} \bibnamefont{and}
  \bibinfo{author}{\bibfnamefont{A.}~\bibnamefont{Boudaoud}},
  \bibinfo{journal}{Physical Review Letters} \textbf{\bibinfo{volume}{96}},
  \bibinfo{pages}{136103} (\bibinfo{year}{2006}).

\bibitem[{\citenamefont{Balankin et~al.}(2010)\citenamefont{Balankin, Ochoa,
  Miguel, Ortiz, and Cruz}}]{PhysRevE.81.061126}
\bibinfo{author}{\bibfnamefont{A.~S.} \bibnamefont{Balankin}},
  \bibinfo{author}{\bibfnamefont{D.~S.} \bibnamefont{Ochoa}},
  \bibinfo{author}{\bibfnamefont{I.~A.} \bibnamefont{Miguel}},
  \bibinfo{author}{\bibfnamefont{J.~P.~n.} \bibnamefont{Ortiz}},
  \bibnamefont{and} \bibinfo{author}{\bibfnamefont{M.~A.~M.}
  \bibnamefont{Cruz}}, \bibinfo{journal}{Physical Review E}
  \textbf{\bibinfo{volume}{81}}, \bibinfo{pages}{061126}
  (\bibinfo{year}{2010}).

\bibitem[{\citenamefont{Lin et~al.}(2009)\citenamefont{Lin, Sun, Hsiao, Hwu,
  Wang, and Hong}}]{lin2009spontaneous}
\bibinfo{author}{\bibfnamefont{Y.-C.} \bibnamefont{Lin}},
  \bibinfo{author}{\bibfnamefont{J.-M.} \bibnamefont{Sun}},
  \bibinfo{author}{\bibfnamefont{J.-H.} \bibnamefont{Hsiao}},
  \bibinfo{author}{\bibfnamefont{Y.}~\bibnamefont{Hwu}},
  \bibinfo{author}{\bibfnamefont{C.}~\bibnamefont{Wang}}, \bibnamefont{and}
  \bibinfo{author}{\bibfnamefont{T.-M.} \bibnamefont{Hong}},
  \bibinfo{journal}{Physical review letters} \textbf{\bibinfo{volume}{103}},
  \bibinfo{pages}{263902} (\bibinfo{year}{2009}).

\bibitem[{\citenamefont{Cambou and Menon}(2011)}]{cambou2011three}
\bibinfo{author}{\bibfnamefont{A.~D.} \bibnamefont{Cambou}} \bibnamefont{and}
  \bibinfo{author}{\bibfnamefont{N.}~\bibnamefont{Menon}},
  \bibinfo{journal}{Proceedings of the National Academy of Sciences}
  \textbf{\bibinfo{volume}{108}}, \bibinfo{pages}{14741}
  (\bibinfo{year}{2011}).

\bibitem[{\citenamefont{Keim et~al.}(2019)\citenamefont{Keim, Paulsen,
  Zeravcic, Sastry, and Nagel}}]{keim2019memory}
\bibinfo{author}{\bibfnamefont{N.~C.} \bibnamefont{Keim}},
  \bibinfo{author}{\bibfnamefont{J.~D.} \bibnamefont{Paulsen}},
  \bibinfo{author}{\bibfnamefont{Z.}~\bibnamefont{Zeravcic}},
  \bibinfo{author}{\bibfnamefont{S.}~\bibnamefont{Sastry}}, \bibnamefont{and}
  \bibinfo{author}{\bibfnamefont{S.~R.} \bibnamefont{Nagel}},
  \bibinfo{journal}{Reviews of Modern Physics} \textbf{\bibinfo{volume}{91}},
  \bibinfo{pages}{035002} (\bibinfo{year}{2019}).

\bibitem[{\citenamefont{Corte et~al.}(2008)\citenamefont{Corte, Chaikin,
  Gollub, and Pine}}]{corte2008random}
\bibinfo{author}{\bibfnamefont{L.}~\bibnamefont{Corte}},
  \bibinfo{author}{\bibfnamefont{P.~M.} \bibnamefont{Chaikin}},
  \bibinfo{author}{\bibfnamefont{J.~P.} \bibnamefont{Gollub}},
  \bibnamefont{and} \bibinfo{author}{\bibfnamefont{D.~J.} \bibnamefont{Pine}},
  \bibinfo{journal}{Nature Physics} \textbf{\bibinfo{volume}{4}},
  \bibinfo{pages}{420} (\bibinfo{year}{2008}).

\bibitem[{\citenamefont{Keim and Nagel}(2011)}]{keim2011generic}
\bibinfo{author}{\bibfnamefont{N.~C.} \bibnamefont{Keim}} \bibnamefont{and}
  \bibinfo{author}{\bibfnamefont{S.~R.} \bibnamefont{Nagel}},
  \bibinfo{journal}{Physical review letters} \textbf{\bibinfo{volume}{107}},
  \bibinfo{pages}{010603} (\bibinfo{year}{2011}).

\bibitem[{\citenamefont{Regev et~al.}(2021)\citenamefont{Regev, Attia, Dahmen,
  Sastry, and Mungan}}]{regev2021topology}
\bibinfo{author}{\bibfnamefont{I.}~\bibnamefont{Regev}},
  \bibinfo{author}{\bibfnamefont{I.}~\bibnamefont{Attia}},
  \bibinfo{author}{\bibfnamefont{K.}~\bibnamefont{Dahmen}},
  \bibinfo{author}{\bibfnamefont{S.}~\bibnamefont{Sastry}}, \bibnamefont{and}
  \bibinfo{author}{\bibfnamefont{M.}~\bibnamefont{Mungan}},
  \bibinfo{journal}{arXiv preprint arXiv:2101.01083}  (\bibinfo{year}{2021}).

\bibitem[{\citenamefont{Mungan et~al.}(2019)\citenamefont{Mungan, Sastry,
  Dahmen, and Regev}}]{mungan2019networks}
\bibinfo{author}{\bibfnamefont{M.}~\bibnamefont{Mungan}},
  \bibinfo{author}{\bibfnamefont{S.}~\bibnamefont{Sastry}},
  \bibinfo{author}{\bibfnamefont{K.}~\bibnamefont{Dahmen}}, \bibnamefont{and}
  \bibinfo{author}{\bibfnamefont{I.}~\bibnamefont{Regev}},
  \bibinfo{journal}{Physical Review Letters} \textbf{\bibinfo{volume}{123}},
  \bibinfo{pages}{178002} (\bibinfo{year}{2019}).

\bibitem[{\citenamefont{Sethna et~al.}(1993)\citenamefont{Sethna, Dahmen,
  Kartha, Krumhansl, Roberts, and Shore}}]{sethna1993hysteresis}
\bibinfo{author}{\bibfnamefont{J.~P.} \bibnamefont{Sethna}},
  \bibinfo{author}{\bibfnamefont{K.}~\bibnamefont{Dahmen}},
  \bibinfo{author}{\bibfnamefont{S.}~\bibnamefont{Kartha}},
  \bibinfo{author}{\bibfnamefont{J.~A.} \bibnamefont{Krumhansl}},
  \bibinfo{author}{\bibfnamefont{B.~W.} \bibnamefont{Roberts}},
  \bibnamefont{and} \bibinfo{author}{\bibfnamefont{J.~D.} \bibnamefont{Shore}},
  \bibinfo{journal}{Physical Review Letters} \textbf{\bibinfo{volume}{70}},
  \bibinfo{pages}{3347} (\bibinfo{year}{1993}).

\bibitem[{\citenamefont{Bense and van Hecke}(2021)}]{bense2021complex}
\bibinfo{author}{\bibfnamefont{H.}~\bibnamefont{Bense}} \bibnamefont{and}
  \bibinfo{author}{\bibfnamefont{M.}~\bibnamefont{van Hecke}},
  \bibinfo{journal}{Proceedings of the National Academy of Sciences}
  \textbf{\bibinfo{volume}{118}} (\bibinfo{year}{2021}).

\bibitem[{\citenamefont{Keim et~al.}(2020)\citenamefont{Keim, Hass, Kroger, and
  Wieker}}]{keim2020global}
\bibinfo{author}{\bibfnamefont{N.~C.} \bibnamefont{Keim}},
  \bibinfo{author}{\bibfnamefont{J.}~\bibnamefont{Hass}},
  \bibinfo{author}{\bibfnamefont{B.}~\bibnamefont{Kroger}}, \bibnamefont{and}
  \bibinfo{author}{\bibfnamefont{D.}~\bibnamefont{Wieker}},
  \bibinfo{journal}{Physical Review Research} \textbf{\bibinfo{volume}{2}},
  \bibinfo{pages}{012004} (\bibinfo{year}{2020}).

\bibitem[{\citenamefont{van Hecke}(2021)}]{van2021profusion}
\bibinfo{author}{\bibfnamefont{M.}~\bibnamefont{van Hecke}},
  \bibinfo{journal}{Physical Review E} \textbf{\bibinfo{volume}{104}},
  \bibinfo{pages}{054608} (\bibinfo{year}{2021}).

\bibitem[{\citenamefont{Lindeman and Nagel}(2021)}]{lindeman2021multiple}
\bibinfo{author}{\bibfnamefont{C.~W.} \bibnamefont{Lindeman}} \bibnamefont{and}
  \bibinfo{author}{\bibfnamefont{S.~R.} \bibnamefont{Nagel}},
  \bibinfo{journal}{Science Advances} \textbf{\bibinfo{volume}{7}},
  \bibinfo{pages}{eabg7133} (\bibinfo{year}{2021}).

\bibitem[{\citenamefont{C. and D.}(2021)}]{keim2021multiperiodic}
\bibinfo{author}{\bibfnamefont{N.}~\bibnamefont{C.}, \bibfnamefont{Keim}}
  \bibnamefont{and} \bibinfo{author}{\bibfnamefont{J.}~\bibnamefont{D.},
  \bibfnamefont{Paulsen}}, \bibinfo{journal}{Science Advances}
  \textbf{\bibinfo{volume}{7}}, \bibinfo{pages}{eabg7685}
  (\bibinfo{year}{2021}).

\bibitem[{\citenamefont{Deutsch et~al.}(2004)\citenamefont{Deutsch, Dhar, and
  Narayan}}]{deutsch2004return}
\bibinfo{author}{\bibfnamefont{J.}~\bibnamefont{Deutsch}},
  \bibinfo{author}{\bibfnamefont{A.}~\bibnamefont{Dhar}}, \bibnamefont{and}
  \bibinfo{author}{\bibfnamefont{O.}~\bibnamefont{Narayan}},
  \bibinfo{journal}{Physical Review Letters} \textbf{\bibinfo{volume}{92}},
  \bibinfo{pages}{227203} (\bibinfo{year}{2004}).

\bibitem[{\citenamefont{Perkovi{\'c} and Sethna}(1997)}]{perkovic1997improved}
\bibinfo{author}{\bibfnamefont{O.}~\bibnamefont{Perkovi{\'c}}}
  \bibnamefont{and} \bibinfo{author}{\bibfnamefont{J.~P.}
  \bibnamefont{Sethna}}, \bibinfo{journal}{Journal of applied physics}
  \textbf{\bibinfo{volume}{81}}, \bibinfo{pages}{1590} (\bibinfo{year}{1997}).

\bibitem[{\citenamefont{Lechenault and
  Adda-Bedia}(2015)}]{lechenault2015generic}
\bibinfo{author}{\bibfnamefont{F.}~\bibnamefont{Lechenault}} \bibnamefont{and}
  \bibinfo{author}{\bibfnamefont{M.}~\bibnamefont{Adda-Bedia}},
  \bibinfo{journal}{Physical review letters} \textbf{\bibinfo{volume}{115}},
  \bibinfo{pages}{235501} (\bibinfo{year}{2015}).

\bibitem[{\citenamefont{Sethna et~al.}(2001)\citenamefont{Sethna, Dahmen, and
  Myers}}]{sethna2001crackling}
\bibinfo{author}{\bibfnamefont{J.~P.} \bibnamefont{Sethna}},
  \bibinfo{author}{\bibfnamefont{K.~A.} \bibnamefont{Dahmen}},
  \bibnamefont{and} \bibinfo{author}{\bibfnamefont{C.~R.} \bibnamefont{Myers}},
  \bibinfo{journal}{Nature} \textbf{\bibinfo{volume}{410}},
  \bibinfo{pages}{242} (\bibinfo{year}{2001}).

\bibitem[{\citenamefont{Plummer and Nelson}(2020)}]{plummer2020buckling}
\bibinfo{author}{\bibfnamefont{A.}~\bibnamefont{Plummer}} \bibnamefont{and}
  \bibinfo{author}{\bibfnamefont{D.~R.} \bibnamefont{Nelson}},
  \bibinfo{journal}{Physical Review E} \textbf{\bibinfo{volume}{102}},
  \bibinfo{pages}{033002} (\bibinfo{year}{2020}).

\bibitem[{\citenamefont{Preisach}(1935)}]{preisach1935magnetische}
\bibinfo{author}{\bibfnamefont{F.}~\bibnamefont{Preisach}},
  \bibinfo{journal}{Zeitschrift f{\"u}r physik} \textbf{\bibinfo{volume}{94}},
  \bibinfo{pages}{277} (\bibinfo{year}{1935}).

\bibitem[{\citenamefont{Mayergoyz and
  Friedman}(1988)}]{mayergoyz1988generalized}
\bibinfo{author}{\bibfnamefont{I.~D.} \bibnamefont{Mayergoyz}}
  \bibnamefont{and} \bibinfo{author}{\bibfnamefont{G.}~\bibnamefont{Friedman}},
  \bibinfo{journal}{IEEE transactions on Magnetics}
  \textbf{\bibinfo{volume}{24}}, \bibinfo{pages}{212} (\bibinfo{year}{1988}).

\bibitem[{\citenamefont{Middleton}(1992)}]{middleton1992asymptotic}
\bibinfo{author}{\bibfnamefont{A.~A.} \bibnamefont{Middleton}},
  \bibinfo{journal}{Physical Review Letters} \textbf{\bibinfo{volume}{68}},
  \bibinfo{pages}{670} (\bibinfo{year}{1992}).

\bibitem[{\citenamefont{Mungan and Terzi}(2019)}]{mungan2019structure}
\bibinfo{author}{\bibfnamefont{M.}~\bibnamefont{Mungan}} \bibnamefont{and}
  \bibinfo{author}{\bibfnamefont{M.~M.} \bibnamefont{Terzi}}, in
  \emph{\bibinfo{booktitle}{Annales Henri Poincar{\'e}}}
  (\bibinfo{organization}{Springer}, \bibinfo{year}{2019}),
  vol.~\bibinfo{volume}{20}, pp. \bibinfo{pages}{2819--2872}.

\bibitem[{\citenamefont{Gilbert et~al.}(2015)\citenamefont{Gilbert, Chern,
  Fore, Lao, Zhang, Nisoli, and Schiffer}}]{gilbert2015direct}
\bibinfo{author}{\bibfnamefont{I.}~\bibnamefont{Gilbert}},
  \bibinfo{author}{\bibfnamefont{G.-W.} \bibnamefont{Chern}},
  \bibinfo{author}{\bibfnamefont{B.}~\bibnamefont{Fore}},
  \bibinfo{author}{\bibfnamefont{Y.}~\bibnamefont{Lao}},
  \bibinfo{author}{\bibfnamefont{S.}~\bibnamefont{Zhang}},
  \bibinfo{author}{\bibfnamefont{C.}~\bibnamefont{Nisoli}}, \bibnamefont{and}
  \bibinfo{author}{\bibfnamefont{P.}~\bibnamefont{Schiffer}},
  \bibinfo{journal}{Physical Review B} \textbf{\bibinfo{volume}{92}},
  \bibinfo{pages}{104417} (\bibinfo{year}{2015}).

\bibitem[{\citenamefont{Kedia et~al.}(2019)\citenamefont{Kedia, Pan, Slotine,
  and England}}]{kedia2019drive}
\bibinfo{author}{\bibfnamefont{H.}~\bibnamefont{Kedia}},
  \bibinfo{author}{\bibfnamefont{D.}~\bibnamefont{Pan}},
  \bibinfo{author}{\bibfnamefont{J.-J.} \bibnamefont{Slotine}},
  \bibnamefont{and} \bibinfo{author}{\bibfnamefont{J.~L.}
  \bibnamefont{England}}, \bibinfo{journal}{arXiv preprint arXiv:1908.09332}
  (\bibinfo{year}{2019}).

\bibitem[{\citenamefont{Benichou and Givli}(2013)}]{benichou2013structures}
\bibinfo{author}{\bibfnamefont{I.}~\bibnamefont{Benichou}} \bibnamefont{and}
  \bibinfo{author}{\bibfnamefont{S.}~\bibnamefont{Givli}},
  \bibinfo{journal}{Journal of the Mechanics and Physics of Solids}
  \textbf{\bibinfo{volume}{61}}, \bibinfo{pages}{94} (\bibinfo{year}{2013}).

\bibitem[{\citenamefont{Moessner and Ramirez}(2006)}]{moessner2006geometrical}
\bibinfo{author}{\bibfnamefont{R.}~\bibnamefont{Moessner}} \bibnamefont{and}
  \bibinfo{author}{\bibfnamefont{A.~P.} \bibnamefont{Ramirez}},
  \bibinfo{journal}{Phys. Today} \textbf{\bibinfo{volume}{59}},
  \bibinfo{pages}{24} (\bibinfo{year}{2006}).

\bibitem[{\citenamefont{Chen et~al.}(2021)\citenamefont{Chen, Pauly, and
  Reis}}]{chen2021reprogrammable}
\bibinfo{author}{\bibfnamefont{T.}~\bibnamefont{Chen}},
  \bibinfo{author}{\bibfnamefont{M.}~\bibnamefont{Pauly}}, \bibnamefont{and}
  \bibinfo{author}{\bibfnamefont{P.~M.} \bibnamefont{Reis}},
  \bibinfo{journal}{Nature} \textbf{\bibinfo{volume}{589}},
  \bibinfo{pages}{386} (\bibinfo{year}{2021}).

\bibitem[{\citenamefont{Jules et~al.}(2022)\citenamefont{Jules, Reid, Daniels,
  Mungan, and Lechenault}}]{jules2022delicate}
\bibinfo{author}{\bibfnamefont{T.}~\bibnamefont{Jules}},
  \bibinfo{author}{\bibfnamefont{A.}~\bibnamefont{Reid}},
  \bibinfo{author}{\bibfnamefont{K.~E.} \bibnamefont{Daniels}},
  \bibinfo{author}{\bibfnamefont{M.}~\bibnamefont{Mungan}}, \bibnamefont{and}
  \bibinfo{author}{\bibfnamefont{F.}~\bibnamefont{Lechenault}},
  \bibinfo{journal}{Physical Review Research} \textbf{\bibinfo{volume}{4}},
  \bibinfo{pages}{013128} (\bibinfo{year}{2022}).

\bibitem[{\citenamefont{Silverberg et~al.}(2014)\citenamefont{Silverberg,
  Evans, McLeod, Hayward, Hull, Santangelo, and Cohen}}]{silverberg2014using}
\bibinfo{author}{\bibfnamefont{J.~L.} \bibnamefont{Silverberg}},
  \bibinfo{author}{\bibfnamefont{A.~A.} \bibnamefont{Evans}},
  \bibinfo{author}{\bibfnamefont{L.}~\bibnamefont{McLeod}},
  \bibinfo{author}{\bibfnamefont{R.~C.} \bibnamefont{Hayward}},
  \bibinfo{author}{\bibfnamefont{T.}~\bibnamefont{Hull}},
  \bibinfo{author}{\bibfnamefont{C.~D.} \bibnamefont{Santangelo}},
  \bibnamefont{and} \bibinfo{author}{\bibfnamefont{I.}~\bibnamefont{Cohen}},
  \bibinfo{journal}{Science} \textbf{\bibinfo{volume}{345}},
  \bibinfo{pages}{647} (\bibinfo{year}{2014}).

\bibitem[{\citenamefont{Udani and Arrieta}(2021)}]{udani2021programmable}
\bibinfo{author}{\bibfnamefont{J.~P.} \bibnamefont{Udani}} \bibnamefont{and}
  \bibinfo{author}{\bibfnamefont{A.~F.} \bibnamefont{Arrieta}},
  \bibinfo{journal}{Extreme Mechanics Letters} \textbf{\bibinfo{volume}{42}},
  \bibinfo{pages}{101081} (\bibinfo{year}{2021}).

\bibitem[{\citenamefont{Waitukaitis et~al.}(2015)\citenamefont{Waitukaitis,
  Menaut, Chen, and Van~Hecke}}]{waitukaitis2015origami}
\bibinfo{author}{\bibfnamefont{S.}~\bibnamefont{Waitukaitis}},
  \bibinfo{author}{\bibfnamefont{R.}~\bibnamefont{Menaut}},
  \bibinfo{author}{\bibfnamefont{B.~G.-g.} \bibnamefont{Chen}},
  \bibnamefont{and}
  \bibinfo{author}{\bibfnamefont{M.}~\bibnamefont{Van~Hecke}},
  \bibinfo{journal}{Physical Review Letters} \textbf{\bibinfo{volume}{114}},
  \bibinfo{pages}{055503} (\bibinfo{year}{2015}).

\bibitem[{\citenamefont{Stern et~al.}(2017)\citenamefont{Stern, Pinson, and
  Murugan}}]{stern2017complexity}
\bibinfo{author}{\bibfnamefont{M.}~\bibnamefont{Stern}},
  \bibinfo{author}{\bibfnamefont{M.~B.} \bibnamefont{Pinson}},
  \bibnamefont{and} \bibinfo{author}{\bibfnamefont{A.}~\bibnamefont{Murugan}},
  \bibinfo{journal}{Physical Review X} \textbf{\bibinfo{volume}{7}},
  \bibinfo{pages}{041070} (\bibinfo{year}{2017}).

\bibitem[{\citenamefont{Chen and Santangelo}(2018)}]{chen2018branches}
\bibinfo{author}{\bibfnamefont{B.~G.-g.} \bibnamefont{Chen}} \bibnamefont{and}
  \bibinfo{author}{\bibfnamefont{C.~D.} \bibnamefont{Santangelo}},
  \bibinfo{journal}{Physical Review X} \textbf{\bibinfo{volume}{8}},
  \bibinfo{pages}{011034} (\bibinfo{year}{2018}).

\bibitem[{\citenamefont{Bitzek et~al.}(2006)\citenamefont{Bitzek, Koskinen,
  G\"ahler, Moseler, and Gumbsch}}]{FIRE}
\bibinfo{author}{\bibfnamefont{E.}~\bibnamefont{Bitzek}},
  \bibinfo{author}{\bibfnamefont{P.}~\bibnamefont{Koskinen}},
  \bibinfo{author}{\bibfnamefont{F.}~\bibnamefont{G\"ahler}},
  \bibinfo{author}{\bibfnamefont{M.}~\bibnamefont{Moseler}}, \bibnamefont{and}
  \bibinfo{author}{\bibfnamefont{P.}~\bibnamefont{Gumbsch}},
  \bibinfo{journal}{Physical Review Letters} \textbf{\bibinfo{volume}{97}},
  \bibinfo{pages}{170201} (\bibinfo{year}{2006}).

\bibitem[{\citenamefont{O’hern et~al.}(2003)\citenamefont{O’hern, Silbert,
  Liu, and Nagel}}]{OHern2003}
\bibinfo{author}{\bibfnamefont{C.~S.} \bibnamefont{O’hern}},
  \bibinfo{author}{\bibfnamefont{L.~E.} \bibnamefont{Silbert}},
  \bibinfo{author}{\bibfnamefont{A.~J.} \bibnamefont{Liu}}, \bibnamefont{and}
  \bibinfo{author}{\bibfnamefont{S.~R.} \bibnamefont{Nagel}},
  \bibinfo{journal}{Physical Review E} \textbf{\bibinfo{volume}{68}},
  \bibinfo{pages}{011306} (\bibinfo{year}{2003}).

\bibitem[{\citenamefont{Phillips}(1979)}]{Phillips1979}
\bibinfo{author}{\bibfnamefont{J.~C.} \bibnamefont{Phillips}},
  \bibinfo{journal}{Journal of Non-Crystalline Solids}
  \textbf{\bibinfo{volume}{34}}, \bibinfo{pages}{153} (\bibinfo{year}{1979}).

\bibitem[{\citenamefont{Alexander}(1998)}]{alexander1998}
\bibinfo{author}{\bibfnamefont{S.}~\bibnamefont{Alexander}},
  \bibinfo{journal}{Physics Reports} \textbf{\bibinfo{volume}{296}},
  \bibinfo{pages}{65} (\bibinfo{year}{1998}).

\bibitem[{\citenamefont{Moukarzel}(1998)}]{moukarzel1998}
\bibinfo{author}{\bibfnamefont{C.~F.} \bibnamefont{Moukarzel}},
  \bibinfo{journal}{Physical Review Letters} \textbf{\bibinfo{volume}{81}},
  \bibinfo{pages}{1634} (\bibinfo{year}{1998}).

\end{thebibliography}

\end{document}